\documentclass{elsart}  
\usepackage{epsfig}
\begin{document}

\begin{frontmatter}

\title{Vlasov stability of the Hamiltonian Mean Field model}

\author{ Celia Anteneodo and Ra\'ul O. Vallejos}

\address{Centro Brasileiro de Pesquisas F\'{\i}sicas,
         R. Dr. Xavier Sigaud 150, \\
         22290-180, Rio de Janeiro, Brazil }

\thanks{e-mail: celia@cbpf.br; vallejos@cbpf.br}

\begin{abstract}
We investigate the dynamical stability of 
a fully-coupled system of $N$ inertial rotators, 
the so-called Hamiltonian Mean Field model.
In the limit $N \to \infty$, and after proper scaling 
of the interactions, the $\mu$-space dynamics is governed 
by a Vlasov equation.  We apply a nonlinear stability test to 
(i) a selected set of spatially homogeneous solutions 
of Vlasov equation, qualitatively similar to those observed
in the quasi-stationary states arising from fully magnetized initial conditions, 
and  (ii) numerical coarse-grained distributions of the finite-$N$ dynamics. 
Our results are consistent with previous numerical evidence 
of the disappearance of the homogenous quasi-stationary family 
below a certain energy.
\end{abstract}

\begin{keyword}   long-range interactions \sep nonextensivity \sep Vlasov equation 
\PACS
05.20.-y         
\sep 05.70.-a    

\end{keyword}

\end{frontmatter}
\vspace*{-4mm}
\section{Introduction}
\label{sec:introduction}
\vspace*{-4mm}
Nonextensive Statistical Mechanics (NSM) was proposed in 1988 \cite{ct88} 
as an attempt to account for many phenomena that can not be described 
within the standard Boltzmann-Gibbs scenario (see \cite{review} for reviews). 
NSM starts from a generalized entropy functional, and using suitable constraints, 
a generalization of the usual canonical ensemble thermostatistics is obtained. 
Instead of the exponential weight, in NSM one has a $q$-exponential probability 
distribution in $\Gamma$-space: 

\begin{equation} \label{qcanonical}
P(x) \propto \exp_q[-\beta H(x)] 
\equiv \left\{ 1- \beta (1-q) H(x) \right\}^{1/(1-q)}      
\end{equation}
(if the quantity between curly brackets is negative then $P=0$).
Here, $x$ is a point in the $N$-body phase space of a system described 
by the Hamiltonian $H(x)$, $\beta$ a generalized inverse temperature and 
$q$ the entropic index \cite{review}. 
Unlike the canonical distribution function [Eq. (\ref{qcanonical}) 
with $q=1$], which applies to equilibrium situations, the $q$-canonical 
distribution (with $q\neq 1$) is believed to be associated to meta-equilibrium 
or even out-of-equilibrium regimes \cite{review}. 

The predictions of the theory can in principle be derived from Eq.~(\ref{qcanonical}). 
However, explicit calculations for the anomalous high-dimensional systems of interest 
for NSM have not been performed yet due to large technical difficulties. 
For instance, single particle distribution functions require the 
integration of $P(x)$ over $N\!-\!1$ particles. 

On the other side, there are many experimental and numerical data well 
described by $q$-exponential functions \cite{review}. 
This agreement is considered as an indirect evidence of the 
applicability of the NSM to those systems. 
Moreover, $q$-exponentials have also been observed in 
non-Hamiltonian and even non-physical (biological, economical, etc.) 
systems \cite{interdisciplinary},  but in these cases the connection with Statistical 
Mechanics (from first principles) is less clear. 

A model system potentially important for NSM is the so-called Hamiltonian 
Mean Field model (HMF), an inertial $XY$ ferromagnet with infinite-range 
interactions \cite{inagaki93,antoni95}:
 
\begin{equation} \label{ham0}
H   =  \frac{1}{2} \sum_{i=1  }^N     p_{i}^{2} +
       \frac{1}{2N} \sum_{i,j=1}^N 
       \left[   1-\cos(\theta_{i}-\theta_{j}) \right] \;, 
\end{equation}
where the coordinates of each particle are the angle $\theta_i$ 
and its angular momentum $p_i$. 
This model can be considered a simple prototype for complex, long-range systems 
like galaxies and plasmas. In a certain sense, the HMF is a descendant of the 
mass-sheet gravitational model \cite{antoni95}.

The HMF has been extensively studied in the 
last few years (see \cite{reviewHMF} for a review), especially due to its long-lived 
quasi-equilibrium states, which exhibit breakdown of ergodicity, 
anomalous diffusion, aging, and non-Maxwell velocity distributions. 
In some cases $q$-exponential behaviors were reported: 
two-time correlation functions \cite{montemurro03,pluchino03},   
single particle velocity distributions (truncated $q$-Gaussians) \cite{qss_vd} and
temperature relaxation \cite{latora04}.
The simplicity of the HMF, that made possible a full analysis of its statistical
properties both in the standard canonical \cite{antoni95} and microcanonical 
ensembles \cite{antoni02}, gives support to the hope that it may be possible 
to derive explicit first principle predictions, e.g., for single 
particle velocity distributions. 

\begin{figure}[htb]
\begin{center}
\includegraphics*[bb=79 349 502 672, width=0.75\textwidth]{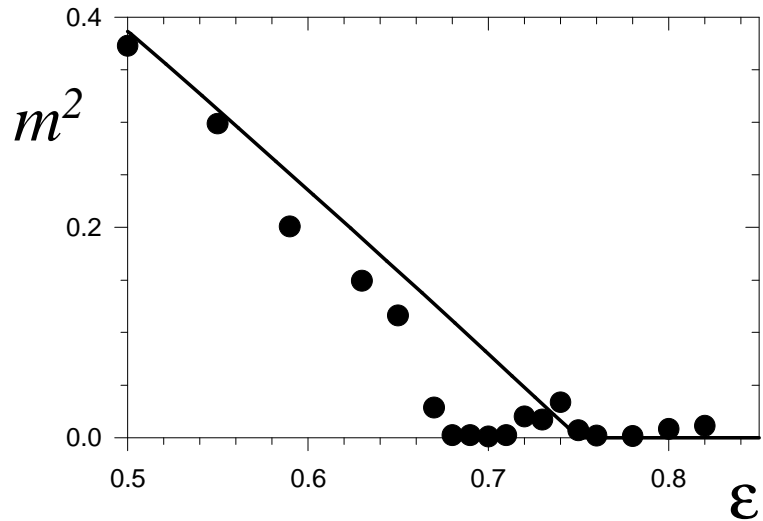}
\end{center}
\caption{%
Squared magnetization versus energy for quasi-stationary states 
(circles). System size is $N=10^4$.
We averaged over 10 initial conditions and in the time window 
$500<t<2000$ (see Sec.~\ref{sec:results} for numerical details).
The full line is the analytical result for thermal equilibrium \cite{antoni95}. 
}
\label{fig:m2vse}
\end{figure}

The most widely investigated quasi-stationary states (QSS) of the HMF arise 
from special initial conditions (water-bag) that evolve to  
states with well defined macroscopic characteristics and whose lifetimes increase   
with the system size \cite{qss_vd}. For a given energy, the QSS is different 
from the corresponding equilibrium state to which the 
system eventually evolves. 
These states have been observed for specific energies 
$\varepsilon\equiv H/N$ below the critical 
value $\varepsilon_{\rm cr}=0.75$ corresponding to a ferromagnetic phase transition. 
In a magnetization diagram (see \cite{qss_vd} and Fig.~\ref{fig:m2vse}), 
a first part of the QSS family 
lies on the continuation of the high-energy equilibrium line (having zero magnetization) 
down to the energy 
$\varepsilon_{\rm min} \simeq 0.68$. 
At $\varepsilon_{\rm min}$ the $m=0$ line ceases to exist, being
substituted by a curve of magnetized states. 

This paper is an attempt to understand the origin of such discontinuity. 
We base our analysis on the associated Vlasov equation which describes 
the continuum limit of the Hamiltonian particle dynamics. 
Using the stability criterion recently developed by Yamaguchi et al. 
\cite{yamaguchi}, we show that the {\em coarse-grained} homogeneous (zero-magnetization)   
QSS family becomes unstable at $\varepsilon_{\rm min}$.  

 \section{Vlasov approach}
 \label{sec:taylor}
\vspace*{-4mm}
Introducing radial and tangent unitary vectors 
$\hat{\bf r}_i =   (\sin \theta_i, \cos \theta_i)$ and 
$\hat{\bf \theta}_i=( \cos \theta_i,-\sin \theta_i)$, 
Hamiltonian (\ref{ham0}) leads to the equations of motion 

\begin{eqnarray} \nonumber
\dot{\theta}_i &=& p_i, \\ \label{motioneqs}
\dot{p}_i      &=& {\bf m} \cdot \hat{\bf \theta}_i,
\end{eqnarray}
for $i=1,\ldots,N$, with the magnetization per particle 
\begin{equation}
{\bf m} =   \frac{1}{N} \sum_{i=  1}^N \hat{\bf r}_i  \; .
\end{equation}
Eqs. (\ref{motioneqs}) describe the two-dimensional motion of a cloud 
of $N$ points in the fluctuating mean-field ${\bf m}(t)$. 
In the continuum approximation, the cloud becomes a fluid governed by the 
Liouville equation for the single-particle probability distribution $f(\theta,p,t)$
\begin{equation} \label{liouville}
\frac{\partial f}{\partial t} \,+\, p\frac{\partial f}{\partial \theta}
\,+\, {\bf m}\cdot \hat{\bf \theta} \; \frac{\partial f}{\partial p}\,=\,0  \;,
\end{equation}
with 
\begin{equation} 
{\bf m}(t)=\int d\theta \,\hat{\bf r}(\theta)\int dp \,f(\theta,p,t) \;.
\end{equation}
Liouville equation (\ref{liouville}) has to be solved self-consistently
with the calculation of ${\bf m}(t)$, and is thus formally equivalent
to the Vlasov-Poisson system found in plasma and gravitational dynamics 
\cite{balescu}. The continuum approximation will be progressively better 
as $N$ grows, especially when the cloud of particles is spread over 
a finite region in $\mu$-space. This is the case of the distributions we will 
analyze in this paper (see Fig.~\ref{fig:pdfs}). Detailed discussions of 
Vlasov approximation can be found in \cite{braun77,spohn80,yamaguchi}.

\section{Dynamical stability}
\label{sec:stability}
\vspace*{-4mm}
Every spatially homogeneous distribution $f(p)$ is a stationary solution of Vlasov 
equation (\ref{liouville}).  The reason is that homogeneity leads to ${\bf m}=0$ and 
thus momenta do not change. The interesting point is whether these homogeneous
solutions are stable or not. There is the well known Landau analysis \cite{balescu}
which concerns {\em linear} stability. This linear criterion was recently applied
to the HMF by Choi and Choi to discuss ensemble equivalence and quasi-stationarity
\cite{choi03}. 
Besides the linearity restriction, Landau analysis has the disadvantage of requiring 
the knowledge of detailed analytical properties of $f(p)$. This excludes, in principle,
numerically obtained momentum distributions.

A more powerful stability criterion for homogeneous equilibria has been proposed by 
Yamaguchi et al. \cite{yamaguchi}. This is a nonlinear criterion specific to the  
HMF. It states that $f(p)$ is stable if and only if the quantity 

\begin{equation}
I\,=\,1\,+\,\frac{1}{2} \int_{-\infty}^\infty {\rm d}p\, \frac{f^\prime(p)}{p}
\end{equation}
is positive (it is assumed that $f$ is an even function of $p$).
Note that this condition is equivalent to the zero frequency case 
of Landau's recipe \cite{antoni95,choi03}. 
 
The authors of Ref.~\cite{yamaguchi} showed that a distribution which is spatially 
homogeneous and Gaussian in momentum becomes unstable below the transition energy
 $\varepsilon_{\rm cr}=3/4$ (see also \cite{inagaki93,choi03}), 
in agreement with analytical and numerical results for finite $N$ systems. 
They also showed that homogeneous states with zero-mean uniform $f(p)$ (water-bag)
are stable above $\varepsilon=7/12=0.58...$ (see also \cite{antoni95,choi03}), 
and verified that the stability of these Vlasov solutions has a counterpart in 
the associated particle dynamics.

\begin{figure}[htb]
\begin{center}
\includegraphics*[bb=20 419 596 701, width=1.00\textwidth]{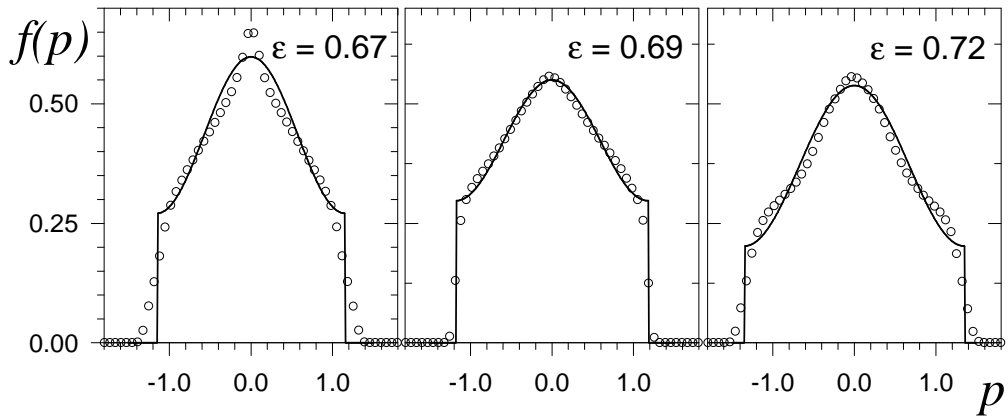}
\end{center}
\caption{Momentum distributions for energies indicated on the figure. 
Histograms (bin size $\simeq$ 0.07) were generated averaging on the time 
interval (500,2000) and 10 different initial conditions 
(very small random perturbations of the fully magnetized regular water-bag). 
In all cases $N=10^4$. 
See Sec.~\ref{sec:results} for further details. 
Full lines are water-bag-plus-cosine fittings to the data.} 
\label{fig:pdfs}
\end{figure}

In the same spirit, it is also worth analyzing the stability of the family 
of $q$-exponentials
\begin{equation} \label{qgaussian}
f(p) \propto \exp_q(-\alpha p^2) =\left[ 1- \alpha (1-q) p^2 \right]^{1/(1-q)} \; ,
\end{equation}
which allow to scan a wide spectrum of probability distributions, from finite-support 
to power-law tailed ones, containing as particular cases the Gaussian ($q=1$) and the
water-bag ($q=-\infty$). In Eq. (\ref{qgaussian}), the normalization constant has 
been omitted and the parameter $\alpha>0$ is related to the second moment 
$\langle p^2 \rangle$, which only exists for $q<5/3$. 
In the homogeneous states of the HMF one has 
$\langle p^2 \rangle = 2\varepsilon-1$, as can be easily derived from Eq. (\ref{ham0}). 
Then, the stability indicator $I$ as a function of the energy for the $q$-exponential 
family reads
\begin{equation}
I = 1-\frac{3-q}{2(5-3q)(2\varepsilon-1) } \;.
\end{equation}
Stability occurs for energies above $\varepsilon_{\rm q}$ 
\begin{equation} \label{threshold}
\varepsilon_{\rm q}(q) = \varepsilon_{\rm cr} +\frac{q-1}{2(5-3q)} \; .
\end{equation}
It is easy to verify that one recovers the known stability thresholds for the 
uniform and Gaussian distributions, i.e., $\varepsilon_{\rm q}(-\infty)=7/12$ and 
$\varepsilon_{\rm q}(1)=3/4$. We remark that Eq.~(\ref{threshold}) states that only 
finite-support distributions,  
corresponding to $q<1$, are stable below $\varepsilon_{\rm cr}$.
This agrees with numerical studies in the QSS regimes of the HMF 
(see Fig.~\ref{fig:pdfs}).

\begin{figure}[htb]
\begin{center}
\includegraphics*[bb=109 266 514 648, width=0.65\textwidth]{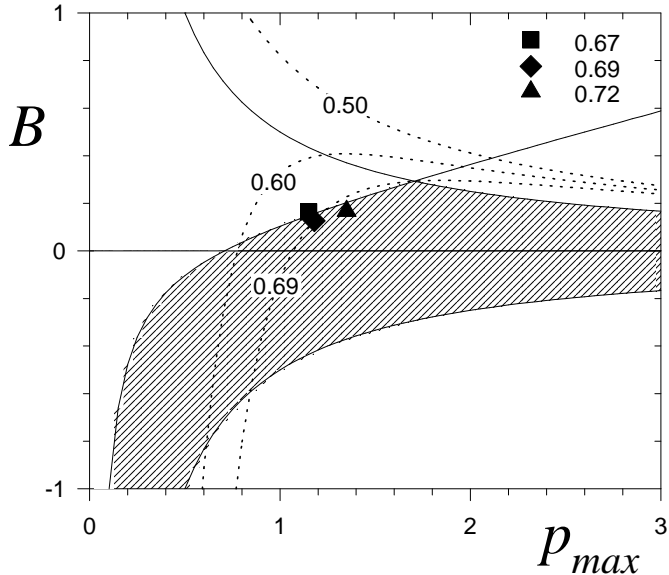}
\end{center}
\caption{Stability diagram for water-bag-plus-cosine momentum 
distributions. The shadowed area corresponds to the stability domain. 
Also shown are three iso-energetic subfamilies with 
$\varepsilon=0.50,0.60,0.69$ (dotted lines). 
Symbols correspond to quasi-stationary states
obtained numerically, for energies $\varepsilon=0.67,0.69,0.72$
and $N=10^4$ (see Sec.~\ref{sec:results} for numerical details). } 
\label{fig:cosine}
\end{figure}

The QSS we discuss in this paper are obtained from particular initial conditions: 
all angles are equal and momenta distributed uniformly with zero mean; energy is
chosen slightly below $\varepsilon_{\rm cr}$.
After an initial ``violent relaxation" (the initial state is highly inhomogeneous)
the system evolves to a QSS  characterized by spatial homogeneity and momentum 
distributions with cutoffs \cite{qss_vd}. However, the resulting
distributions of momenta are not uniform any more. Some structure appears on top 
of a slightly smoothed water-bag (see Fig.~\ref{fig:pdfs}). 
A very simple family of functions exhibiting the basic features of the observed 
$f(p)$ is the water-bag plus cosine:
\begin{equation}
f(p)=
\left\{ \matrix{  1/(2p_{\rm max}) + B \cos(\pi p/p_{\rm max}),  
&  \mbox{for $p \in [-p_{\rm max},p_{\rm max}]$} \;  \cr
                  0,                          
&  \mbox{otherwise}  \; .
  } \right. \end{equation}
The stable functions $f(p)$ occupy a region in parameter space $(p_{\rm max}, B)$
defined by three boundary curves (see Fig.~\ref{fig:cosine}). 
The left-top boundary is given by $I=0$ and the other two are associated to 
positivity, i.e., $f(p)=0$.
The metastable states obtained from numerical simulations of finite 
systems (to be described below) can be approximately fitted by 
water-bag-plus-cosine functions. 
Remarkably, the resulting pairs $(p_{\rm max},B)$ fall close to the 
boundary of Vlasov stability, and exit the stability region for energies 
below the limiting value $\varepsilon \simeq 0.68$.

Fig.~\ref{fig:cosine} also informs that for low energies, 
$\varepsilon<7/12$, there is a qualitative change in the shape 
of stable homogeneous distributions: they change the maximum at $p=0$
by a minimum since $B<0$.
As $\varepsilon\to 1/2$, the stable distribution resembles a couple of delta functions 
approaching the origin.

\section{Numerical results}
\label{sec:results}
\vspace*{-4mm}
The application of the stability criterion to a discrete distribution requires 
some coarse-graining. 
We will consider smoothed distributions 
\begin{equation}
f_\gamma(p)\,=\,\frac{1}{N} \sum_{i=1}^N \delta_\gamma(p-p_i),
\end{equation}
with Lorentzian delta functions

\begin{equation}
\delta_\gamma(x) \;=\; \frac{\gamma}{\pi} \frac{1}{x^2+\gamma^2} .
\end{equation}
The coarse-graining parameter $\gamma$ corresponds to an intermediate scale,   
between microscopic and macroscopic (to be specified). 
We will also impose parity, $f(p)=f(-p)$, so that our smoothed distribution 
will be calculated as 

\begin{equation} \label{fgamma} 
f_\gamma(p)\,=\,\frac{\gamma}{\pi N} \sum_{p_i>0} \left[ 
\frac{1}{ (p-p_i)^2 +\gamma^2}  \,+\, \frac{1}{ (p+p_i)^2 +\gamma^2}  
  \right] \;.
\end{equation}
Accordingly, the stability indicator becomes simply 

\begin{equation} \label{indicator}
I\,=\,1\,+\,\frac{1}{N} \sum_{p_i>0} \frac{p_i^2-\gamma^2}{ (p_i^2 +\gamma^2)^2} \;.
\end{equation}

All simulations have been done with $N=10^4$ particles.
We have considered water-bag initial conditions.
All particles are set at $\theta=0$, corresponding to maximum magnetization
modulus $m=1$. Momenta were distributed on a regular lattice, symmetrically around 
$p=0$. Our experience indicates that this ``crystalline" distribution behaves 
like a larger system, say $N=10^5$, where the $p$'s are extracted randomly
from a uniform distribution. Initial states were propagated according to 
Hamilton equations, approximated by a symplectic fourth-order scheme 
\cite{yoshida90}. 
We registered the quantities of interest $m(t)$ (homogeneity
indicator) and $p_i(t)$ (input to the stability indicator). 
Strictly speaking, $m=0$ does not imply that the states 
are homogeneous. However, the sudden relaxation that leads to the 
present QSS mixes particles  \cite{pluchino03} in such a way 
that $m=0$ and spatial homogeneity are expected to be synonymous.

\begin{figure}[htb]
\begin{center}
\includegraphics*[bb=82 366 508 674, width=0.75\textwidth]{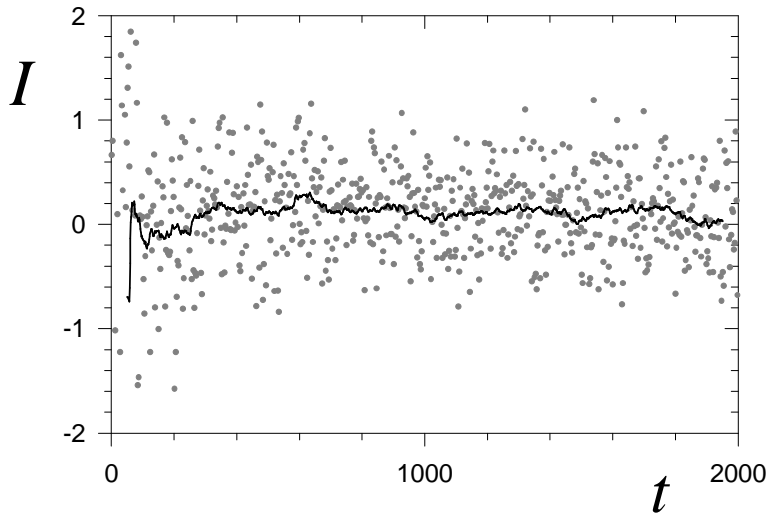}
\end{center}
\caption{Example of stability as a function of time, 
for the fully magnetized regular water-bag initial condition, 
with $\varepsilon=0.69$, $N=10^4$, and smoothing $\gamma=0.05$. 
The full line corresponds to a running average over a time window
of width $\Delta t=100$, indicating quasi-stationarity 
in the interval $500<t<2000$. } 
\label{fig:ivst}
\end{figure}

\begin{figure}[htb]
\begin{center}
\includegraphics*[bb=86 336 522 662, width=0.75\textwidth]{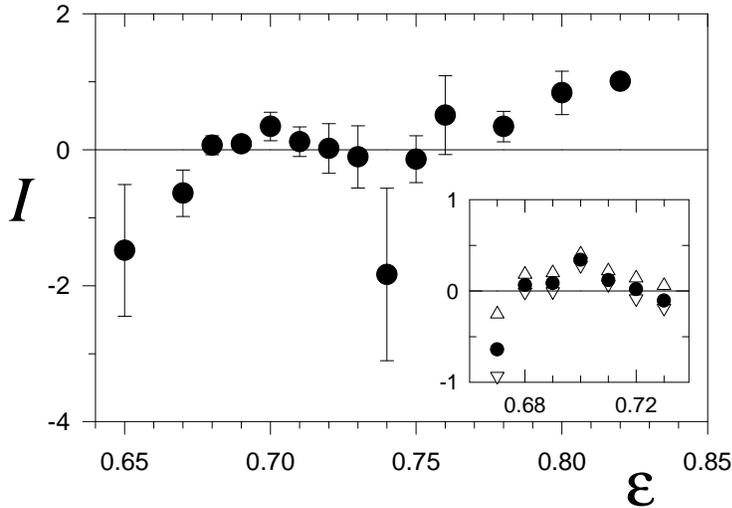}
\end{center}
\caption{Stability versus energy for quasi-stationary states.
Note that $I$ goes through zero around $\varepsilon \simeq 0.68$.
The smoothing parameter is $\gamma=0.05$ ($N=10^4$).
We averaged over 10 initial conditions and in the time interval $500<t<2000$. 
Inset: $I$ vs $\varepsilon$ for $\gamma=0.025$ ($\nabla$), $0.05$ ($\bullet$) and 
$0.1$ ($\triangle$).
}  
\label{fig:ivse}
\end{figure}

Concerning the calculation of smoothed distributions,
we chose a coarse-graining parameter $\gamma=0.05$, after
verifying that such a choice erases microscopic fluctuations, but preserves
macroscopic features of the QSS momentum distributions (like those shown in
Fig.~\ref{fig:pdfs}). Anyway, some tests were run with $\gamma=0.025$ and 
$\gamma=0.1$ to check that our results are not substantially changed 
by the precise choice of the smoothing (see inset in Fig.~\ref{fig:ivse}).
We also eliminated rapid time fluctuations by doing an additional average in 
the time window $500<t<2000$, a typical domain 
of quasi-stationarity after initial relaxation of the water-bag 
(see Fig.~\ref{fig:ivst}).

The sensitivity to initial conditions (of the microscopic state of the system) 
after times like those considered here ($t \approx 1000$) is enormous.  
To be sure that the effect we are observing is statistically robust, 
we also did an average on initial conditions. These were obtained by 
slightly perturbing the regular water-bag with a small random component:
$p_i'=p_i (1+10^{-4} \xi)$, where $\xi$ is random uniform in $[-1,1]$. 
Our stability analysis of the coarse-grained quasi-stationary states of
the HMF is summarized in Fig.~\ref{fig:ivse}. 
There one sees that stability is positive for energies above $\varepsilon\simeq 0.68$, 
except for the small region $0.73 < \varepsilon < 0.75$. We believe that this is
probably a spurious effect due to the proximity of the ferromagnetic phase transition;
further analyses, involving longer times, or larger system sizes, are necessary for
elucidating this question.
The fact that the stability indicator becomes negative below $\varepsilon\simeq 0.68$ 
signals the disappearance of the homogeneous metastable phase at that energy.
Simultaneously with $I$ crossing zero at that point, the magnetization rises
from negligible values to finite ones, of the order of those 
corresponding to equilibrium (Fig.~\ref{fig:m2vse}). This stability
picture is consistent with the analysis of the water-bag-plus-cosine
family (Sec.~\ref{sec:stability}), showing that distributions like 
those observed in QSS live close to a stability border,
and that this border is traversed when energy goes down a limiting value.
The symbols in Fig.~\ref{fig:cosine} were obtained by fitting a water-bag-plus-cosine
to the numerical histograms of Fig.~\ref{fig:pdfs}.

A comment about negative stability is in order.  
The present stability test only applies to homogeneous states. 
However, below $\varepsilon=0.68$ the measured distributions 
are evidently inhomogeneous ($m \neq 0$). In these cases, negative
stability refers to hypothetical homogeneous states having the measured $f(p)$. 

\section{Conclusions}
\label{sec:conclusions}
\vspace*{-4mm}
Yamaguchi et al. conjectured that stable homogeneous solutions of Vlasov equation 
lead, via discretization, to quasi-stationary states of Hamiltonian dynamics, 
and showed that it is indeed the case for some families of states of the HMF \cite{yamaguchi}. 
We have shown that the reciprocal may be also true: metastable states of the discrete 
Hamiltonian dynamics lead, after smoothing, to stable solutions of the Vlasov 
equation (this was verified for the homogeneous QSS).

Moreover, our analysis indicates that the disappearance of the homogenous family 
(resulting from the violent relaxation of special water-bag initial conditions)
at $\varepsilon \simeq 0.68$ can be interpreted as a reflection of the lose of 
Vlasov stability at that critical energy. This interpretation is supported by
similar evidence resulting from the analytical study of the water-bag-plus-cosine
family.

Despite providing interesting information, the results we presented are not definitive. 
Some points remain unclear and deserve further investigation, e.g., 
the stability properties close to the ferromagnetic transition. 
Another point, perhaps more important, concerns the quantification of spatial 
homogeneity. A null squared magnetization indicates the absence of 
spatial structures on the largest scale. A more stringent test would require 
the verification that higher-order Fourier coefficients of the particle density 
tend to zero in the large $N$ limit.

\section*{Acknowledgements}
\vspace*{-4mm}
We are grateful to Constantino Tsallis for the very enthusiastic 
discussions which stimulated this and many other works. 
We thank T. Dauxois and S. Ruffo for interesting conversations
and for communicating their results prior to publication (thanks also
to J. Barr\'e, F. Bouchet and Y. Yamaguchi).
This work had financial support from Brazilian Agencies CNPq, FAPERJ 
and PRONEX.


\end{document}